\documentclass[aps,prb,twocolumn,amsmath,showpacs,showkeys,superscriptaddress,floatfix]{revtex4}

\setcitestyle{numbers,square}

\usepackage{amsmath}
\usepackage{amssymb}
\usepackage{multirow,bigdelim}
\usepackage{array}
\usepackage{cancel}
\usepackage{epsfig}
\usepackage{graphicx}
\usepackage{graphics}
\usepackage{url}
\usepackage[english]{babel}
\usepackage{dcolumn}
\usepackage{color}
\usepackage[squaren]{SIunits}
\usepackage{bbold}
\usepackage[T1]{fontenc}
\usepackage{float}

\newcommand{\Onlinecite}{\cite}
\newcommand{\HHdown}{\raisebox{-1mm}{\rotatebox{90}{$\Lleftarrow$}}}
\newcommand{\HHup}{\raisebox{-1mm}{\rotatebox{90}{$\Rrightarrow$}}}
\newcommand{\LHdown}{\raisebox{-1mm}{\rotatebox{90}{$\Leftarrow$}}}
\newcommand{\LHup}{\raisebox{-1mm}{\rotatebox{90}{$\Rightarrow$}}}
\newcommand{\edown}{\raisebox{-1mm}{\rotatebox{90}{$\leftarrow$}}}
\newcommand{\eup}{\raisebox{-1mm}{\rotatebox{90}{$\rightarrow$}}}
\begin{document}

\title{Influence of quantum dot geometry on p-shell transitions in differently charged quantum dots}

\author{M.~Holtkemper}
\affiliation{Institut f\"ur Festk\"orpertheorie, Universit\"at M\"unster,
Wilhelm-Klemm-Str.~10, 48149 M\"unster, Germany}

\author{D.~E.~Reiter}
\affiliation{Institut f\"ur Festk\"orpertheorie, Universit\"at M\"unster,
Wilhelm-Klemm-Str.~10, 48149 M\"unster, Germany}

\author{T.~Kuhn}
\affiliation{Institut f\"ur Festk\"orpertheorie, Universit\"at M\"unster,
Wilhelm-Klemm-Str.~10, 48149 M\"unster, Germany}

\date{\today}

\begin{abstract}
Absorption spectra of neutral, negatively and positively charged semiconductor quantum dots are studied theoretically. We provide an overview of the main energetic structure around the p-shell transitions, including the influence of nearby nominally dark states.
Based on the envelope function approximation, we treat the four-band Luttinger theory as well as the direct and short range exchange Coulomb interactions within a configuration interaction approach. The quantum dot confinement is approximated by an anisotropic harmonic potential.
We present a detailed investigation of state mixing and correlations mediated by the individual interactions. Differences and similarities between the differently charged quantum dots are highlighted. Especially large differences between negatively and positively charged quantum dots become evident. 
We present a visualisation of energetic shifts and state mixtures due to changes in size, in-plane asymmetry and aspect ratio. Thereby we provide a better understanding of the experimentally hard to access question of quantum dot geometry effects in general. Our findings show a new method to determine the in-plane asymmetry from photoluminescense excitation spectra. Furthermore, we supply basic knowledge for tailoring the strength of certain state mixtures or the energetic order of particular excited states via changes in the shape of the quantum dot, which is highly interesting e.g. to understand relaxation paths.
\end{abstract}


\maketitle

\section{Introduction}\label{sec:Introduction}
Self-assembled semiconductor quantum dots (QDs) confine electronic states within a nanometer size scale, leading to a discrete energetic level structure. Most studies focus on the QD ground states in the prospect of several possible applications within the fields of quantum computing \cite{kok2007linear}, advanced photon sources \cite{michler2000quantum, stevenson2006semiconductor, pan2012multiphoton} and spintronics \cite{wolf2001spintronics}. Less is known about the higher excited level structure, which is vital for the understanding of time-resolved phenomena like relaxation and dephasing mechanisms \cite{htoon2001carrier,huneke2011role,TODOQuelleKonstanz}, recombinations of multiexcitons with more than two electrons or holes \cite{hawrylak1999excitonic,akimov2002fine,chithrani2005electronic,babinski2006emission,arashida2011four,kuklinski2011tuning,molas2012fine,piketka2013photon,molas2016quadexciton} or resonant absorption characteristics, typically measured via photoluminescence excitation (PLE) spectroscopy \cite{gammon1996fine,htoon2001carrier,benny2012excitation,molas2016energy}.

\begin{figure}
\includegraphics[width=0.45\textwidth]{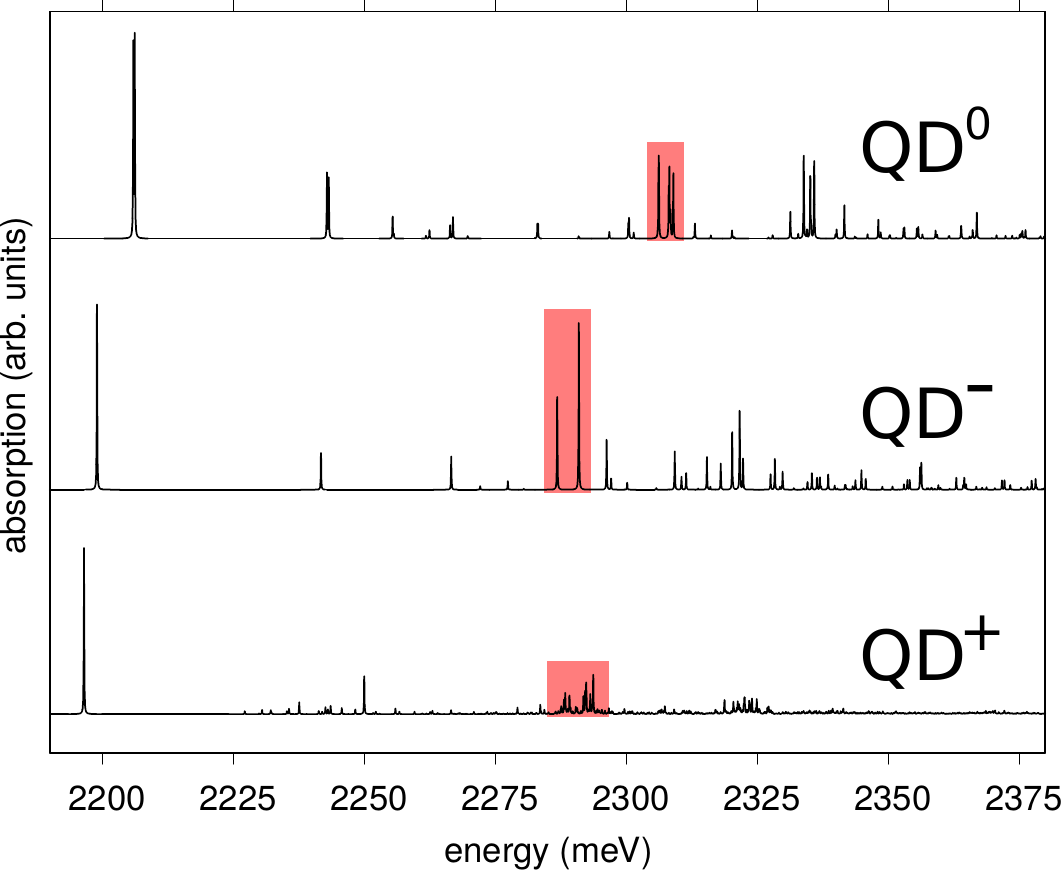}
\caption{(Color online) Typical energy and absorption spectra for single optical excitations in QD$^0$, QD$^-$ and QD$^+$. p-shell transitions are marked by a red background.}\label{fig:overview}
\end{figure}

In contrast to the well separated ground state transition, for the higher excited states a complex spectrum appears depending on the individal QD, in particular on the QD charge and geometry. In Fig.~\ref{fig:overview} we plot typical calculated absorption spectra for neutral (QD$^0$), negatively (QD$^-$) and positively (QD$^+$) charged QDs. Our focus is on transitions between the first excited heavy hole and electron states, which are called p-shell transitions (marked in Fig.~\ref{fig:overview} by a red background).

In this paper we provide a theoretical analysis to explain the differences and similarities in the QD spectra.
Therefore we will address three main questions:
Firstly, we ask which states can be expected around the p-shell transitions and how do they influence each other.
Secondly, we want to know which fundamental differences and similarities between QD$^0$, QD$^-$ and QD$^+$ exist.
Thirdly, we ask how the energetic structure, the spin and the spatial contributions of the different lines change within different QD geometries.
To answer these questions, we study the influence of the Coulomb interaction, including direct and short range exchange contributions, as well as valence band mixing effects via a Luttinger model. Thereby we show that correlations play a vital role.

This article is structured as follows. After the introduction in section~\ref{sec:Introduction}, containing a brief overview of existing studies, we discuss our model in section~\ref{sec:Qdmodel}. In section~\ref{sec:Interactions}, the different interactions are studied one by one. Thereby the influence of the correlations and the differences and similarities between the differently charged QDs are clarified. In section~\ref{sec:Geometry}, the QD size, asymmetry and aspect ratio are altered and trends in absorption spectra are discussed. A conclusion is given in section~\ref{sec:Conclusion}.

\subsection{Overview of existing studies}\label{sec:Introduction:overview}
Although an all-embracing study of excited states in different QD geometries is still missing (to the best of our knowledge), there have been several studies highlighting particular aspects:

Some studies consider excited states, but are restricted to exemplary, fixed QD geometries. Those studies revealed the typical fine structure of the p-shell transitions in QD$^-$ \cite{ware2005polarized}, QD$^0$ \cite{smolenski2016fine} and QD$^+$ \cite{molas2016excited}. Neighboring transitions are often neglected, although they can be strongly mixed with the p-shell transitions. Such a state mixture changes the behavior of the p-shell transitions, e.g. their spin composition and their relaxation / dephasing properties. Besides, the neighboring states become optically addressable themselves. A larger number of those nearby states is discussed e.g. in Refs. \Onlinecite{gammon1996fine,bester2003pseudopotential,ware2005polarization,glazov2007fine,zielinski2010atomistic,jovanov2011direct,benny2012excitation,molas2016energy}.

To generalize those findings of exemplary QD's, one has to study spectra for different QD geometries systematically, often done via variations in QD size, in-plane asymmetry or aspect ratio. Most of the following studies are either restricted to the easy to access ground states or theoretical works without a rigorous treatment of the crucial Coulomb interaction. Nevertheless, those studies provide fundamental knowledge, that is helpful for our considerations, thus we will review them in the following. Additionally, basic theoretical works provide some coupling strength scaling of the Coulomb interaction.

\underline{Size:} A reduction in QD size will increase the energies of all electronic states \cite{brus1984size,schmidt1986quantum}. The rough energetic level structure is studied in Refs. \Onlinecite{ekimov1993absorption, norris1994measurement, norris1996measurement, laheld1997excitons}. Coulomb coupling strength scalings were revealed in Refs. \Onlinecite{takagahara1993effects, takagahara2000theory}. Coulomb exchange interaction coupling strengths grow around an order of magnitude in QDs compared to their bulk values \cite{woggon1997exchange}. Studies focussed on QD size effects in QD$^0$ consider the stokes shift of the ground states \cite{nirmal1995observation,efros1996band,norris1996size,chamarro1996enhancement,woggon1997exchange,bayer2002fine} or the splitting of the lowest bright doublet due to fine structure splitting (FSS) in Refs. \Onlinecite{bayer2002fine,seguin2005size,kindel2010exciton,tong2011theory,huo2014volume}.

\underline{Asymmetry:} Considering the in-plane asymmetry of QDs, the energy dependency of the hole states was studied in Refs. \Onlinecite{kumar2006theoretical,kumar2013effect}. Effects on FSS (via Coulomb exchange interaction coupling strength) are studied in Refs. \Onlinecite{takagahara1993effects, goupalov1998anisotropic, takagahara2000theory, seguin2005size, kindel2010exciton, tong2011theory}. Highly interesting findings considering our problem are given in Ref. \Onlinecite{hawrylak2010theory}, where an overview of energy shifts in QD$^0$ is given, including excited states and a rigorous treatment of Coulomb interactions.

\underline{Aspect ratio:} Considering different QD aspect ratios, calculated one particle energies of strained pyramidial QDs have been compared in Ref. \Onlinecite{kim1998comparison}. Experimental data about the shell like character of the energetic structure are provided in Ref. \Onlinecite{kuklinski2011tuning}.

\section{QD Model}\label{sec:Qdmodel}
Our calculations are based on a configuration interaction approach within an envelope function approximation. Our Hamiltonian consist of the effective mass energies (EMA) including the QD confinement, the direct (DCI) and short range exchange (SRE) Coulomb interactions as well as the valence band mixing via the offdiagonal elements of a four-band Luttinger model (LUT): $$\hat{H} = \hat{H}_\text{EMA} + \hat{H}_\text{DCI} + \hat{H}_\text{SRE} + \hat{H}_\text{LUT} \, .$$ All calculations are applied to single optical excitations in single self-assembled CdSe QDs. In the following we will explain the four parts of our Hamiltonian in detail.

\subsection{EMA including QD confinement}\label{sec:Qdmodel:confinement}
In the envelope function approximation, the single particle wave functions $\Psi_{a,b}(\vec{r})=\sqrt{V_\text{uc}}\,\Phi_a(\vec{r})\,u_b(\vec{r})$ are separated into a mesoscopic (envelope) part $\Phi_a$ and a microscopic part $u_b$ varying within a unit cell of volume $V_\text{uc}$. The index $b$ in the microscopic part denotes the band index and spin quantum numbers. We consider heavy hole (HH) and light hole (LH) bands as well as the lowest electron conduction bands (EL) with their angular momentum projections $\pm\frac{3}{2}$ (\HHup, \HHdown), $\pm\frac{1}{2}$ (\LHup, \LHdown) and $\pm\frac{1}{2}$ (\eup, \edown), respectively. The relative phase between these states is given in appendix~\ref{sec:phase}.
The envelope functions are defined by a parabolic QD confinement, treated in Cartesian coordinates. Hence we expand the envelope functions in terms of Cartesian harmonic oscillator (HO) eigenfunctions with quantum numbers $a=(a_x,a_y,a_z)$. Due to the emerging shell-like structure, QDs are often called artificial atoms and the naming convention of atomic shells is imitated. Here, states with increasing $a_x+a_y+a_z=0,1,2...$ are called s, p, d ...-like states. LH states are labeled by capital letters S, P, D ... . If necessary, indices label the direction of the excitation, e.g. $d_{xy}$ for $a=(1,1,0)$. One should note, that there exists a similar but different notation for a treatment in spherical coordinates \cite{ekimov1993absorption}. 
To fix the width of the HO confinement, we will specify the QD diameters $L_{\alpha}$ ($\alpha\in \{ x,y,z\}$), that are related to the HO confinement via the frequency $\omega_{b,\alpha}=\frac{4 \hbar}{m_{b,\alpha} \beta^2_{b} L^2_{\alpha}}$. Thus $L_{\alpha}$ denotes those points, where the electron ground state probability density is reduced to $\frac{1}{e}$ of its maximum. This choice is arbitrary, limiting our model to statements about relative changes in QD size. The effective masses $m_{b,\alpha}$ can be deduced from the Luttinger parameters \footnote{$m_\text{HH}^{x/y}=\frac{m_0}{\gamma_1+\gamma_2}$, $m_\text{HH}^{z}=\frac{m_0}{\gamma_1-2\gamma_2}$, $m_\text{LH}^{x/y}=\frac{m_0}{\gamma_1-\gamma_2}$, $m_\text{LH}^{z}=\frac{m_0}{\gamma_1+2\gamma_2}$}. For CdSe \cite{ekimov1993absorption, laheld1997excitons} we use $m_\text{EL}=0.13m_0$ and $\gamma_1=2.1$, $\gamma_2=\gamma_3=0.55$, which leads to $m_\text{HH}^{x/y}\approx 0.38m_0$, $m_\text{HH}^{z}=m_0$, $m_\text{LH}^{x/y}\approx 0.65m_0$ and $m_\text{LH}^{z}\approx 0.31m_0$.
To account for different confinement lengths between electron and hole states \footnote{The energetic difference between the conduction (valence) band edge of the QD and the surrounding material defines the strength of the confinement potential of the QD. These energy differences are not equal for the conduction and valence bands, leading to a different confinement for electrons and holes.}, the QD diameter of the holes is multiplied by a factor $\beta_\text{HH}=\beta_\text{LH}=\beta$. Here a value of $\beta = 1.15$ is used. We set $\beta_\text{EL} = 1$.
In addition to the effective mass energies within the HO confinement, we add half of the bandgap (1840 meV for CdSe \cite{ekimov1993absorption,laheld1997excitons}) to each of the particle energies.

Interactions between different particles are treated within a configuration interaction approach. Therefore a proper choice of basis states is necessary. A reasonable convergence is achieved by considering single particle states up to $a=(5,5,5)$ for holes and $a=(3,3,3)$ for electrons and antisymmetric combinations of the single particle states up to a weighted index sum $\sum_\text{particles} a_x+a_y+2\cdot a_z$ of 7.

We should state that the HO-confinement preserves inversion symmetry, thus there is no mixing between states with even/odd parity.

\subsection{Coulomb interactions}\label{sec:Qdmodel:coulomb}
The Coulomb interaction matrix elements are described including DCI and SRE, following Refs. \Onlinecite{takagahara1993effects,takagahara1999dephasing,takagahara2000theory,hawrylak2010theory}. The DCI is described by 
\begin{align*}
V^\text{DCI}=&\delta_{b_1,b_4} \delta_{b_2,b_3} \frac{e^2}{4\pi\epsilon_0 \epsilon_r} \cdot \notag\\
& \int d^3R\int d^3R' \frac{\Phi_{a_1}^*(\vec{R})\Phi_{a_2}^*(\vec{R}')\Phi_{a_3}(\vec{R}')\Phi_{a_4}(\vec{R})}{|\vec{R}-\vec{R}'|} .
\end{align*}
Here a screening by the static dielectric constant of the bulk material $\epsilon_r=9.2$ for CdSe \cite{laheld1997excitons} is assumed. The integrals in $V^\text{DCI}$ are evaluated in Fourier space.

SRE is described by the matrix elements
\begin{align*}
V^\text{SRE}= M T \int d^3R \, \Phi_{a_1}^*(\vec{R})\Phi_{a_2}^*(\vec{R})\Phi_{a_3}(\vec{R})\Phi_{a_4}(\vec{R})
\end{align*}
with
\begin{align*}
T = 
\bordermatrix{
 &\HHdown \edown &\LHdown \edown & \LHup \edown  & \HHup \edown &\HHdown \eup &\LHdown \eup & \LHup \eup  & \HHup \eup    \cr
 & 0 &       &           &      &      &           &           &         \cr
 &   &\frac{1}{3}&       &      &\frac{1}{\sqrt{3}}& &         &         \cr
 &   &       &\frac{2}{3}&      &      &\frac{2}{3}&           &         \cr
 &   &       &           &  1   &      &           &\frac{1}{\sqrt{3}}&  \cr
 &   &\frac{1}{\sqrt{3}}& &     &  1   &           &           &         \cr
 &   &       &\frac{2}{3}&      &      &\frac{2}{3}&           &         \cr
 &   &     & &\frac{1}{\sqrt{3}}&      &           &\frac{1}{3}&         \cr
 &   &       &           &      &      &           &           &    0    \cr
}.
\end{align*}
The parameter $M$ is fixed at $M=576$ meV$\cdot$nm$^3$ in CdSe by fitting experimental data from Ref. \Onlinecite{TODOQuelleKonstanz}. A more convenient parameter $M_\text{SRE}$, with the unit of an energy, is defined by $M=M_\text{SRE} \frac{3\pi}{4} a_\text{Bohr}^3$ with the bulk Bohr radius $a_\text{Bohr}$ in CdSe of around 5.5 nm \cite{norris1996measurement,fu1999excitonic}. Thereby we get $M_\text{SRE} \approx 1.47$~meV.

The long range Coulomb exchange interaction (see Refs. \Onlinecite{gammon1996fine,goupalov1998anisotropic,takagahara2000theory}) and higher order terms of SRE $\sim \vec{J}^3 \vec{S}$ (see Refs. \Onlinecite{puls1999magneto, kesteren1990fine, bayer2002fine}), both typically associated with FSS, are neglected in this study.

\subsection{Valence band mixing}\label{sec:Qdmodel:valenceband}
The off-diagonal elements of the Luttinger operator (diagonal elements are already included via EMA) are described in Ref. \Onlinecite{chow1994} via
\begin{align*}
V^\text{LUT} =  
\bordermatrix{
 &\HHdown &\LHdown & \LHup  & \HHup    \cr
 &        & -b     & -c     &          \cr
 & -b^{*} &        &        & -c       \cr
 & -c^{*} &        &        &  b       \cr
 &        & -c^{*} &  b^{*} &          \cr
}
\end{align*}
\begin{align*}
\text{with} \quad b = & \frac{\sqrt{3} \hbar^{2}}{m_{0}} \gamma_{3} \, k^{z}_{\Phi} (k^{x}_{\Phi} - i \, k^{y}_{\Phi} ) \notag\\
c = & \frac{\sqrt{3} \hbar^{2}}{2 m_{0}} (\gamma_{2} ( k^{x}_{\Phi} \, k^{x}_{\Phi} - k^{y}_{\Phi} \, k^{y}_{\Phi} ) - 2 i \gamma_{3} \, k^{x}_{\Phi} \, k^{y}_{\Phi}    )
\end{align*}
with the operator $k_\Phi^x =-i \frac{\partial}{\partial x}$ (analog for $k_\Phi^y$ and $k_\Phi^z$) acting only on the envelope functions. The matrix elements in the used HO basis are known analytically.

In this model we neglect strain effects. Strain is typically present in QDs, but strongly depends on the fabrication parameters. Thus for its inclusion, we would have to assume QD-specific parameters, that would distract from the influence of the QD geometry. However, effects of strain would influence the valence band mixing (that is qualitative already included via LUT) or cause a relative energetic shift of the light hole states, whose energetic positions are therefore badly described in our model and thus typically not further discussed, but whose inclusion is important to mediate some interactions between the heavy hole states. Neglect of strain and the simple harmonic approximation to the confinement allow us to use a large configuration interaction basis and therefore to perform an accurate investigation of the correlations induced by the Coulomb interactions, enabling a good comparability between differently charged QDs. Another approximation in this model is the neglect of piezoelectric effects (also related to strain), that are typically associated with FSS \cite{seguin2005size}, as the long range Coulomb exchange interaction. Quantitative statements about the FSS are thus beyond the scope of our model.

\subsection{Absorption spectra}\label{sec:Qdmodel:absorption}
To calculate the linear absorption spectra, we use the dipole matrix elements
\begin{align*}
\vec{P}\sim \int d^3R \, \Phi^*_{a_e}(\vec{R}) \Phi_{a_h}(\vec{R}) \cdot \vec{\mu}_{b_e, b_h}
\end{align*}
with
\begin{align*}
\vec{\mu}_{b_e, b_h} = \bordermatrix{
        & &\HHdown &\LHdown & \LHup  & \HHup    \cr
 \edown & & 0 & \sqrt{\frac{1}{3}}\vec{e}_{\sigma^-} & \sqrt{\frac{2}{3}}\vec{e}_{\pi^z} & \vec{e}_{\sigma^+} \cr
 \eup   & & \vec{e}_{\sigma^-} & \sqrt{\frac{2}{3}}\vec{e}_{\pi^z} & \sqrt{\frac{1}{3}}\vec{e}_{\sigma^+} & 0 \cr
}.
\end{align*}
For $\beta=1$, we get $\int d^3R \, \Phi^*_{a_e}(\vec{R}) \Phi_{a_h}(\vec{R}) = \delta_{a_e,a_h}$, thus only transitions between the same envelope states are allowed.
For charged QDs, we assume a uniform initial distribution of both spin orientations of the single electron ($0^-$) / heavy hole ($0^+$) ground states.
Thus most relevant trion states are composed of this ground state access carrier and an additional optically created exciton. 
When we label electronic states, we will list the hole states, followed by the electron states. For example a trion consisting of $p_y$ HH and $s$ and $p_x$ electrons will be described by $p_y -s p_x$ in the following. When we want to address all exciton and trion states associated to the same transition, e.g. $p_y-p_x$, $p_y-sp_x$ and $sp_y-p_x$, we will call them $p_y\to p_x$.
Without LUT, all transitions are unpolarized, whereas with LUT they become linearly polarized in QD$^0$ and elliptically polarized in charged QDs (considering a fully aligned spin of the initial particle). In this study we focus on the energetic positions of the different transitions. To be able to see all existing bright states, we study the absorption of a circularly polarized electric field $\vec{E}\sim \vec{e}_{\sigma^-}$. It should be mentioned that also the elliptically polarized transitions in charged QDs have the same absorption amplitude for $\sigma^+$ and $\sigma^-$ polarized light, caused by the supposed uniform initial distribution of the ground state spins. The absorption $\alpha$ is obtained via Fermi's golden rule $\alpha(\omega)\sim |\vec{e}_{\sigma^-}\cdot \langle X^c|\hat{\vec{P}}| 0^c \rangle|^2 \cdot \delta(E_{X^c}-E_{0^c}-\hbar \omega)$ with $c\in \{ 0,-,+\}$ and the exciton / trion states $X^c$. For a better visibility, we widened each peak in the absorption spectra by a Lorentzian function with a full width at half maximum of $0.1$~meV.

\section{Interactions}\label{sec:Interactions}

In this section, we analyze the effects of the different interactions by switching them on successively. The QD shape is fixed at a flat and slightly asymmetric geometry of ($5.8$, $5.0$, $2.0$) nm$^3$.

\subsection{Basic shifts and splittings}\label{sec:Interactions:basicsplittings}

\begin{figure*}
\includegraphics[width=1.0\textwidth]{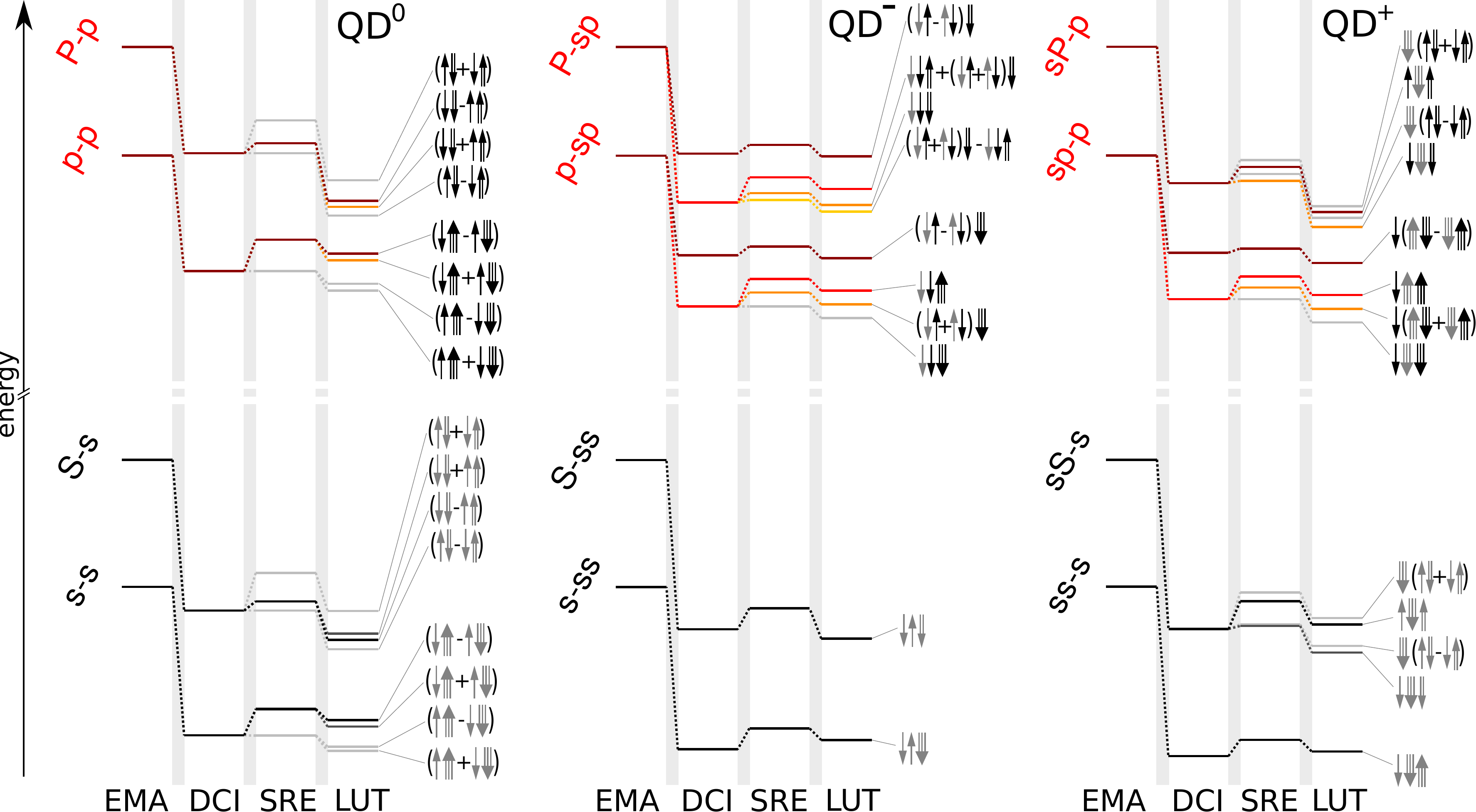}
\caption{(Color online) Schematic plot of energetic shifts and splittings caused by different interactions for exemplary states in QD$^0$, QD$^-$ and QD$^+$. Parameters for a CdSe QD with a shape of ($5.8$, $5.0$, $2.0$) nm$^3$ are used. Dark states (under the assumption of a polarization $\sim \vec{e}_{\sigma}$) are displayed by light grey lines. The main spin contributions are displayed by arrows at the right side of each diagram. Grey (black) arrows are used for particles in a $s$ ($p$) envelope state. In charged QDs, just one spin combination of the Kramers doublet is indicated.}\label{fig:INT}
\end{figure*}

To get a fundamental idea of the energetic structure within QDs, we will consider the different spin configurations of the most prominent transitions between envelope states, namely the $s\to s$, $S\to s$, $p\to p$ and $P\to p$ transitions. Shifts and splittings of the associated electronic states due to DCI, SRE and LUT are shown for QD$^0$, QD$^-$ and QD$^+$ in Fig.~\ref{fig:INT}.

In QD$^0$, there are $2^2=4$ possible combinations of the two EL and two HH / LH spin projections to excitons, respectively. The initially four fold degenerate HH / LH exciton levels are shown on the left hand side of Fig.~\ref{fig:INT}. By turning on DCI (first grey box), all energies are shifted several tens of meV towards lower energies, giving the main contribution to the exciton binding energy. Within the HH excitons ($s-s$ and $p-p$), SRE introduces a splitting of several meV into a dark and a bright doublet at lower / higher energy, as well as small couplings between HH and LH with the same total spin projection. A further small splitting of the doublets of typically several hundreds of $\mu$eV as well as small perturbations of the labeled spin projections arise due to a combination of SRE and LUT. In QDs with cylindrical symmetry (cylindrical QD confinement and strain distribution) the bright doublet remains degenerate. Within the LH excitons ($S-s$ and $P-p$), SRE introduces a splitting into two single states and a doublet in between. Thereby the lowest single state is dark, the doublet is bright and the highest single state is dark in our case, but would become bright under an excitation with an electric field polarized in $z$ direction. The remaining degeneracy of the doublet is then lifted by the combination of SRE and LUT (in QDs with broken cylindrical symmetry).

In QD$^-$, the trions have a half integer total spin. Thus, without a magnetic field, we expect at least a degeneracy of two of all energetic levels reflecting Kramers theorem. If both electrons are in the same envelope state ($s-ss$ and $S-ss$), their spin is fixed to an antiparallel projection and just the spin of the hole can vary. Thus two possible trion spin combinations, or one Kramers doublet, occurs. This doublet is red shifted several tens of meV due to DCI. SRE and LUT just cause smaller disturbances to this energy and spin contributions.
If the electrons are in different envelope states ($p-sp$ and $P-sp$), the three particles of the trion can form $2^3=8$ possible spin combinations. In this case DCI introduces a binding energy of several tens of meV and additionally an electron-electron-exchange term, that causes a strong singlet-triplet splitting (here around 15 meV). This exchange is much stronger than the exchange via SRE between particles in different bands (also see Ref. \Onlinecite{akimov2005electron}). SRE causes a further separation of the triplets (comparable with the bright-dark splitting in QD$^0$, thus around several meV) as well as a small perturbation of the exact separation into singlet and triplet (hole spin disturbs the electron spin alignment). In HH trions ($p-sp$), the lowest triplet is dark, whereas the others are bright. In LH trions ($P-sp$), all triplets are bright. LUT introduces energetic shifts and deviations of the labeled spin combinations.

In QD$^+$, the fundamental shifts and splittings of trions containing two HH ($ss-s$ and $sp-p$) are the same as for the corresponding trions in QD$^-$ ($s-ss$ and $p-sp$), which are also associated to the $s\to s$ and $p\to p$ transitions. A special case occurs in positive trions consisting of EL + HH + LH ($sS-s$ and $sP-p$). In those states there is no singlet-triplet splitting via DCI, because all particles are in different bands. Here SRE introduces a separation into four doublets. The energetically lowest and next to last doublets are bright, whereas the others are dark in our case, but would become bright by using a $\vec{e}_z$ polarization.

There are two relevant conclusions which we obtain from this analysis:

1. In this simplified picture, there are no differences between HH transitions in QD$^-$ and QD$^+$ (see also Ref. \Onlinecite{kavokin2003fine}).

2. The energetic shifts of the different interactions differ by one order of magnitude. This allows us to study the effects of DCI ($\sim$ 10 meV), SRE ($\sim$ 1 meV) and LUT ($\sim$ 0.1 meV) in the following successively, while the weaker interaction typically does not change the overall findings of the stronger one.

Though we presented typical shifts and splittings in our full model, we skipped an explicit consideration of nearby envelope states, i.e. the mixing of states introduced by the respective interactions. Such a reduced description of QD states is widely used, because firstly, the above described basic level structure is in good agreement with findings concerning energetically well separated envelope states (like $s\to s$ states) and secondly, one can use simple theoretical models \cite{bayer2002fine}, just considering the few possible spin states and an effective treatment of the interactions to achieve the above energetic level structure.
However, when envelope states come energetically close, the interplay between them can become important. They can become strongly mixed, which leads to significant changes in optical activity, problems with the assigning of the states (e.g. whether they are more like $p\to p$ triplet or $g\to s$ singlet states), strong deviations of the above described energetic level structure or larger deviations from the labeled spin contributions. In the following, we will analyze these mixtures between different envelope states in detail and clarify their importance. We will do this by a detailed study of the absorption spectra, while turning on the different interactions.

\subsection{QD confinement}\label{sec:Interactions:confinement}

To start with, we look at the absorption spectra in Fig.~\ref{fig:DCI} that are labeled by $\frac{1}{\epsilon_r}=0$. In these spectra just EMA energies are considered, whereas DCI, SRE and LUT are neglected. The main transitions are between holes and electrons that are both in the same envelope state, namely transitions from HH to EL both in the ground state ($s\to s$, black solid line) or both in the first excited states ($p_x\to p_x$, $p_y\to p_y$, reddish solid lines; $p_z \to p_z$ are energetically far above due to the flat QD shape) and from LH to electron both in the ground state ($S\to s$, black dashed line). Therefore we see four corresponding lines in the associated spectra.

With $\beta\neq 1$, also transitions with an even envelope quantum number difference in each direction ($|a^{\text{hole}}_{\alpha} - a^{\text{electron}}_{\alpha}| \in \{0, 2, 4, ... \}$ with $\alpha\in \{x, y, z\}$) are allowed in principle, but their oscillator strength is negligible here. In the considered energetic range mainly $d_{xx}\to s$, $d_{yy}\to s$, $d_{zz}\to s$ (blue solid lines), $g_{xxxx}\to s$, $g_{xxyy}\to s$, $g_{yyyy}\to s$ (green lines), LH $D_{xx}\to s$, $D_{yy}\to s$ (blue dashed lines) and  LH $G_{xxxx}\to s$, $G_{xxyy}\to s$, $G_{yyyy}\to s$ (green dashed lines) transitions are technically allowed.

Without further interactions between the particles, it does not make a difference, if there is an additional carrier in the QD or not. Thus the absorption lines in the differently charged QDs just differ by the oscillator strengths of some peaks, caused by different degeneracies of the associated states.

We remark, that we find highly excited HH $g\to s$ transitions energetically close to $p\to p$ transitions. This becomes obvious, when we recall that the in-plane effective mass of the HH is approximately three times larger than the effective mass of the electron. Thus an increase of the electron in-plane quantum number $a^\text{electron}_{x/y}$ by one needs about the same energy as an increase of the HH in-plane quantum number $a^\text{HH}_{x/y}$ by three.

\subsection{DCI}\label{sec:Interactions:DCI}

\begin{figure}
\includegraphics[width=0.45\textwidth]{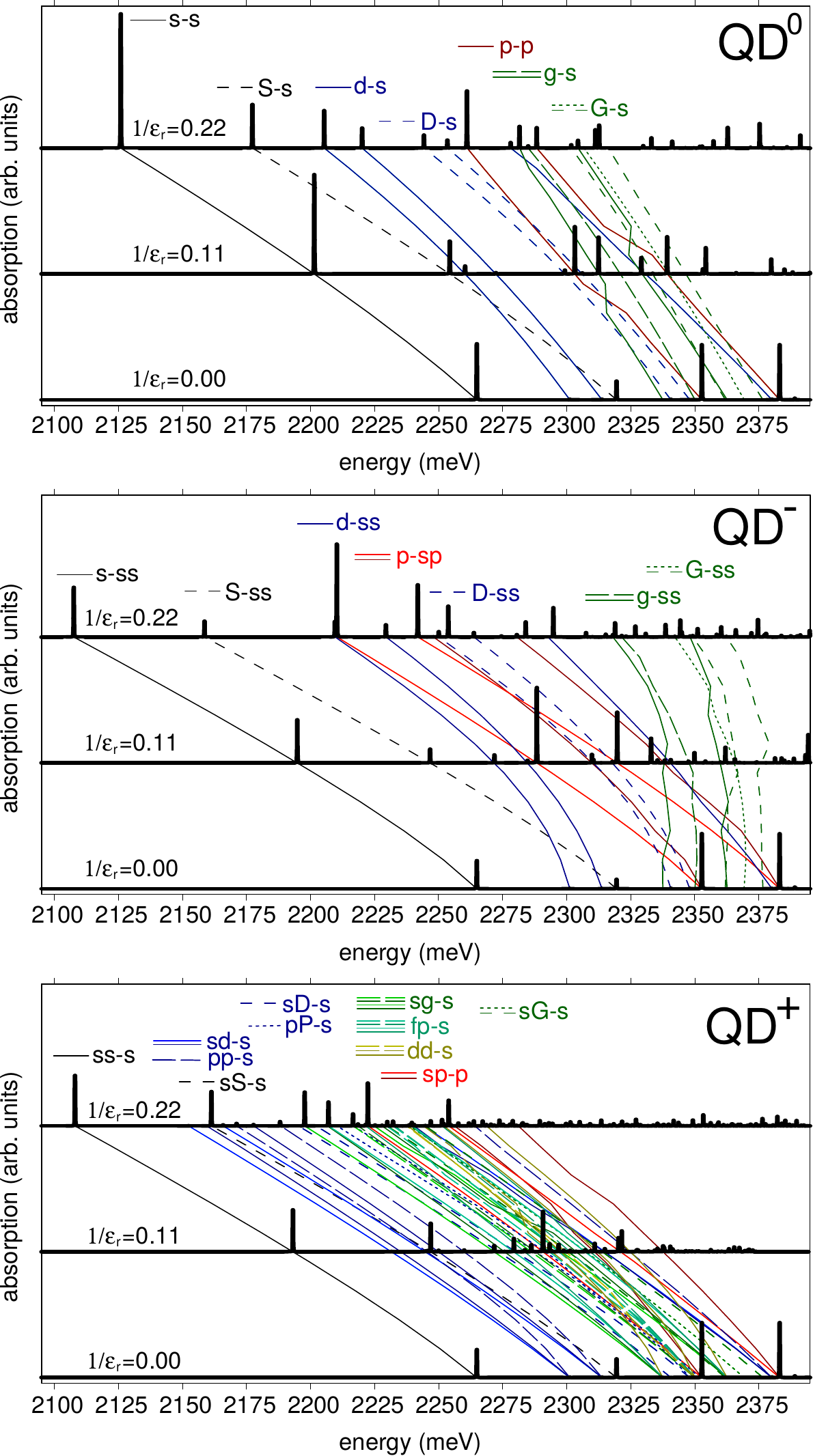}
\caption{(Color online) Absorption spectra labeled with $\frac{1}{\epsilon_r}=0$ belong to calculations in EMA without further interactions. Other spectra take additionally DCI into account, at which an increasing $\frac{1}{\epsilon_r}$ correspond to an increasing DCI coupling strength. Colored lines serve as a guide to the eye.}\label{fig:DCI}
\end{figure}

In the following, we analyze the influence of DCI in detail. To this extend, we plot the absorption spectra for increasing coupling strengths of DCI ($\sim \frac{1}{\epsilon_r}$) in Fig.~\ref{fig:DCI}.
We should mention, that the following variation in DCI coupling strength is not just a theoretical illustration, but could be achieved experimentally by a change of the QD material or, without extensive side effects as changes in effective masses, by a change of the QD size (see section~\ref{sec:Geometry:Volume}).

In Fig.~\ref{fig:DCI}, we plot for clarity the spectra for the values $\frac{1}{\epsilon_r}= 0.00$, $0.11$ and $0.22$, while for intermediate values the energetic positions of the transitions are shown by the colored auxiliary lines. A realistic value for CdSe is $\frac{1}{\epsilon_r}\approx 0.11$. At the positions of anticrossings (e.g. between $p_x-p_x$ and $g_{xxxx}-s$ in QD$^0$ at $\frac{1}{\epsilon_r}\approx 0.08$), the particular auxiliary lines cross instead of merge into each other to further improve clarity. One can identify five main features with increasing DCI:

1. As expected from our basic considerations, a red shift with higher DCI coupling strength occurs. Additionally, a singlet-triplet splitting of the $p\to p$ states arises just in charged QDs (reddish lines).

2. Some peaks seem to appear with increasing DCI coupling strength, clearly visible at e.g. the first two blue labeled lines in QD$^0$ or QD$^-$. With DCI, state mixtures are introduced between states fulfilling the quantum number differences $\Delta (\sum_{\text{particle}} a^{\text{particle}}_{\alpha}) \in \{0, 2, 4, ... \}$. The most relevant admixtures are those of the bright $s\to s$, $p\to p$ into some nearly dark states reached by $d\to s$, $g\to s$, $D\to s$ or $G\to s$ transitions, leading to partially highly increased oscillator strengths of those states, that have been optically allowed in principle already without DCI due to $\beta\neq 1$. We should state that not the dipole transition probability, like with $\beta\neq1$, but the states themselves are changed here. Fortunately these state admixtures are often sufficiently small, thus we will still label our peaks by the mainly contributing one particle combination. Next to the increasing DCI coupling strength, the extent of these mixtures strongly depends on the energetic distance to their coupling partners (see e.g. crossing of $p_x-p_x$ and $g_{xxxx}-s$ in QD$^0$ at $\frac{1}{\epsilon_r}\approx 0.08$ with highly increased amplitude of the $g_{xxxx}-s$ transition).

3. The intensity of the $s\to s$ peaks increase with higher DCI coupling strength. This is due to constructive correlations with higher optically allowed states. The same effect occurs in $p\to p$ states, but losses of intensity to nearby states (as described in the previous paragraph) often compensate the increase.

4. There are much more states around the $p\to p$ transitions in QD$^+$ than in the other QDs.
This has two reasons: Considering $g\to s$ transitions, we have four energy quanta in the $g$-like hole. In QD$^0$ and QD$^-$ these energy quanta are attributed to the one hole (the energy quanta can be disposed in the different in-plane directions to $g_{xxxx}$, $g_{xxxy}$, $g_{xxyy}$, $g_{xyyy}$ or $g_{yyyy}$; 3 of these are bright already, 2 become slightly bright with LUT) and we get five possible transitions. In QD$^+$, we can separate the four energy quanta on the two holes ($g$+$s$, $f$+$p$ or $d$+$d$, all strongly coupled, each with different $x$ and $y$ combinations), giving 19 possible transitions. It should be noticed, that $fp-s$ trions that contain no HH ground state become bright due to admixtures of $sg-s$ and thereby bright $sp-p$. Similar effects occur in nearby transitions like $d\to s$. Because the hole energies are much lower than the electron energies, these effects occur in QD$^-$ only for very high energies (e.g. $s\to g$ transitions are far above the considered energies). Secondly, there are more possible spin combinations in $g\to s$ states with an additional ground state hole than with an additional ground state electron. This is because the two holes in QD$^+$ are in different envelope states enabling eight possible trion-spin combinations, whereas the two electrons in QD$^-$ are in the same envelope state, just enabling two trion-spin combinations. Concerning DCI, we therefore also observe a singlet-triplet splitting of the $sd-s$, $sg-s$, $fp-s$ and LH $sD-s$ states in QD$^+$, but not in the corresponding states in QD$^-$.

5. Each state shifts in energy, when the DCI coupling strength is changed. These shifts are typically different, if we compare either different envelope states in one QD or the same envelope state in differently charged QDs. To understand the underlying mechanisms of these shifts in detail, one should consider the different transitions separately:

\underline{$s\to s$ ($p\to p$):} To understand the different binding energies of $s\to s$ ($p\to p$ analog) states, we have to consider two mechanisms (also see Ref. \Onlinecite{rodt2005correlation}).
\begin{figure}
\includegraphics[width=0.2\textwidth]{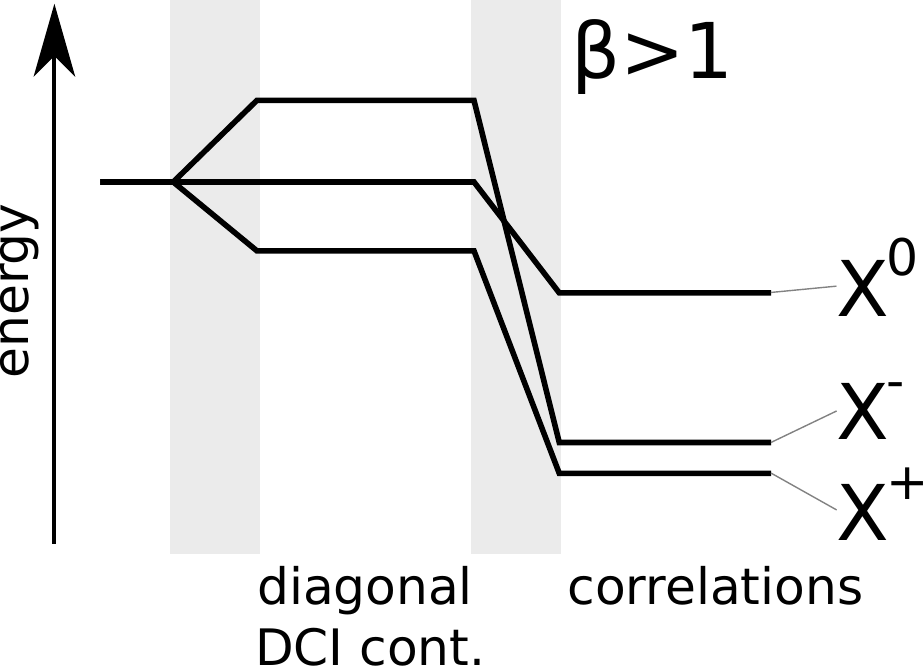}
\caption{Sketch of $s\to s$ ($p\to p$) binding energy contributions for $\beta > 1$.}\label{fig:GSbinding}
\end{figure}
Both lead to energy shifts of the same order of magnitude:
Firstly, there are different contributions of the diagonal DCI elements. In all QDs we have at least an exciton with a certain electron-hole-binding energy. In charged QDs, the additional carrier introduces an additional electron-hole-binding but also a repulsion between the doubly occurring carriers. These additional binding and repulsion do not cancel completely, because the confinement length of holes and electrons is typically different, leading to larger contributions of diagonal DCI elements for the stronger confined particle. In our case ($\beta>1$) we obtain $|V_\text{hhhh}|<|V_\text{heeh}|<|V_\text{eeee}|$, thus in QD$^+$ the additional electron-hole-binding overweights the hole-hole-repulsion and in QD$^-$ the electron-electron-repulsion overweights the additional binding, leading to $E^{s\to s}(\text{QD$^+$}) < E^{s\to s}(\text{QD$^0$}) < E^{s\to s}(\text{QD$^-$})$ (see Fig.~\ref{fig:GSbinding}).
Secondly, correlations to higher states shift all considered states to lower energies. In charged QDs, there are more coupling partners due to more possible spin combinations, leading to a larger red shift than in QD$^0$. Caused by $\beta>1$ the correlations are smaller in QD$^+$ than in QD$^-$, because the coupling strengths are larger between stronger confined states. Thus the correlations shift the relative energetic position of the QD$^-$ ground state towards lower energies, possibly lower than the QD$^0$ ground state or even the QD$^+$ ground state. For $\beta<1$ the role of QD$^+$ and QD$^-$ in this argumentation would switch.
In our explicit calculations for CdSe (Fig.~\ref{fig:DCI}), we see a similar energetic position of the charged QDs, both noticeably below the QD$^0$ ground state. Similar energies were measured in Refs. \Onlinecite{leger2007valence,kuklinski2011tuning,kazimierczuk2011magnetophotoluminescence} for CdTe QDs. For $p\to p$ transitions, the same mechanisms are important. Caused by stronger correlations, the energetic distance between charged QD states and QD$^0$ states is larger than for $s\to s$ transitions (in agreement with experimental results in Ref. \Onlinecite{molas2016energy}) and QD$^-$ states are energetically lower than QD$^+$ states.

\underline{$d\to s$ ($g\to s$):}  Considering $d\to s$ ($g\to s$ analog) transitions, we observe the energetic order $$E^{d\to s}(\text{QD$^+$}) < E^{d\to s}(\text{QD$^0$}) < E^{d\to s}(\text{QD$^-$}).$$ This can be understood by comparing the diagonal energy contributions of DCI. With a much stronger coupling between two particles within the same envelope state, compared with a coupling between two different envelope states ($V^\text{ss}>V^\text{ds}$), we get the energy contributions for the differently charged QDs via: {\small{$$\underbrace{-V_\text{heeh}^\text{ds}-(V_\text{heeh}^\text{ss}-V_\text{hhhh}^\text{ds})}_{\text{QD$^+$}} < \underbrace{-V_\text{heeh}^\text{ds}}_{\text{QD$^0$}} < \underbrace{-V_\text{heeh}^\text{ds}+(V_\text{eeee}^\text{ss}-V_\text{heeh}^\text{ds})}_{\text{QD$^-$}}$$}}
In other words, the two repulsive electrons in QD$^-$ are in the same envelope state, leading to a reduced binding energy in contrast to QD$^+$, where the two repulsive holes are in different envelope states. A special trend occurs in QD$^-$ $d-ss$ transitions, that have a strong bending due to the strongly increasing correlation with the energetically fast approaching $p-sp$ with larger DCI (see Fig.~\ref{fig:DCI}).

Putting the different shifts of $s\to s$, $p\to p$ and $d\to s$, $g\to s$ together, we find for the surrounding of the $p_x\to p_x$ transitions in the present QD geometry and material the following:
In QD$^-$, $p-sp$ states have a much stronger binding energy than $d-ss$ and $g-ss$ states, what means they shift faster towards lower energies with increasing DCI. Thus just for very small DCI, $g-ss$ become important (energetically close and strongly correlated with the bright $p-sp$ states). $d-ss$ become important for a relatively small interval of higher DCI coupling strength. In large regions of DCI coupling strength, $p-sp$ stays energetically clear cut and weakly correlated to other states.
In QD$^+$, $sd-s$ and $sg-s$ have a much stronger binding energy than in QD$^-$, similar to the binding energy of $sp-p$ states. Thus we observe $sg-s$ to be energetically close and strongly mixed with $sp-p$ over the hole range of DCI coupling strength.
In QD$^0$ we have an intermediate situation. $g-s$ slowly shift away from $p-p$ with increasing DCI.

\subsection{SRE}\label{sec:Interactions:SRE}

\begin{figure}
\includegraphics[width=0.45\textwidth]{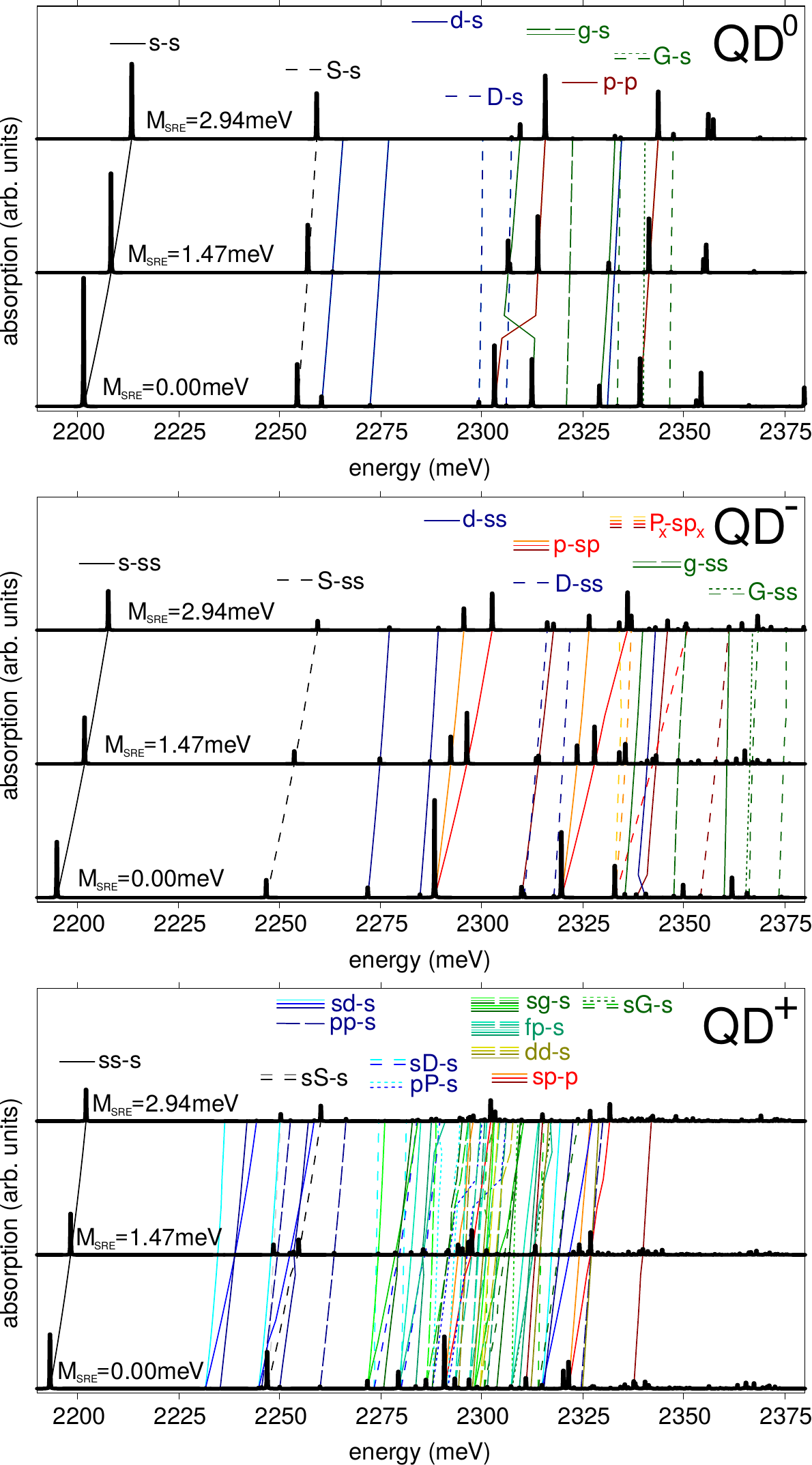}
\caption{(Color online) Absorption spectra in EMA + DCI and different coupling strengths of SRE (spectra labeled by $M_\text{SRE}$ in meV). Colored lines serve as a guide to the eye.}\label{fig:SRE}
\end{figure}

Now we analyze the effects of SRE. Therefore we fix DCI at an appropriate value for CdSe ($\epsilon_r=9.2$, see section~\ref{sec:Qdmodel}) and turn on SRE by increasing $M_\text{SRE}$ in Fig.~\ref{fig:SRE}. Thereby mainly different spin states within the multiplets become separated, as discussed in Fig.~\ref{fig:INT}. Related to the bright states visible in absorption (see Fig.~\ref{fig:SRE}), an overall shift to higher energies can be observed, independently on the QD charge.

Perceivable correlation effects due to SRE are rare. The symmetry concerning quantum number differences of possible correlations is equal to the one described with DCI, thus the smaller effects of SRE are overlain. But SRE also mixes \textit{different} spin configurations with the same total spin, here bright LH and HH states. Due to the weak interaction strength, this is just important when interacting states would (anti-)cross in their energetic positions. In our case this is visible for example in QD$^0$ between $D_y -s$ and $p_x -p_x$, $g_x -s$ around $M_\text{SRE}\approx 1.7$ meV (look closely at right blue dashed line). There the otherwise negligible oscillator strength of the $D_y -s$ transition increases drastically, caused by mixing effects with the bright $p_x -p_x$, $g_x -s$ state.

\subsection{Luttinger}\label{sec:Interactions:Luttinger}

\begin{figure}
\includegraphics[width=0.45\textwidth]{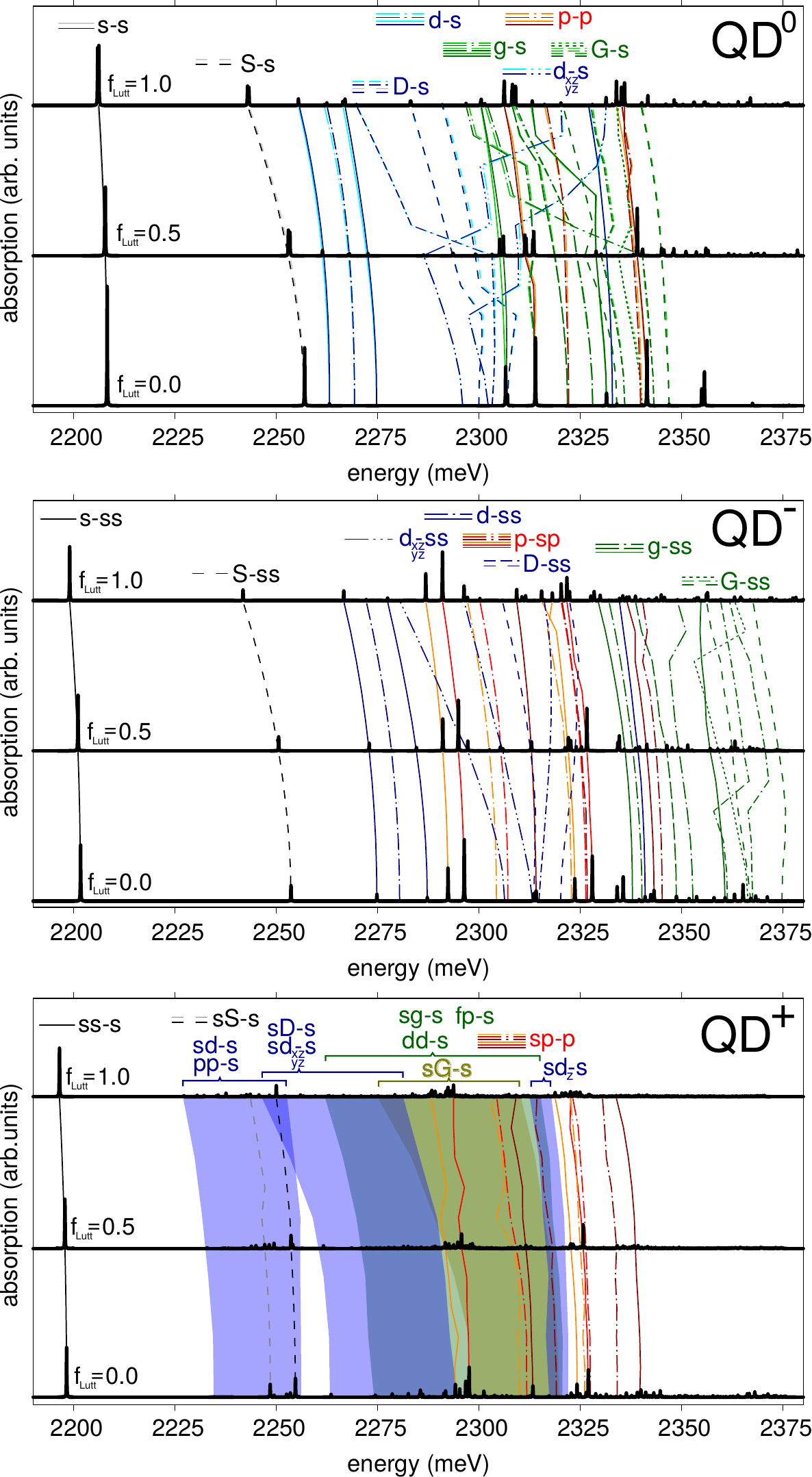}
\caption{(Color online) Absorption spectra in EMA + DCI + SRE and different coupling strengths of LUT. Colored lines serve as a guide to the eye. Lines containing doted sections label transitions, that become bright due to LUT. In QD$^+$, colored areas label the average position of important groups of peaks.}\label{fig:LUT}
\end{figure}

Finally we discuss the effects of LUT. Therefore we fix DCI and SRE at appropriate values for CdSe ($\epsilon_r=9.2$, $M_\text{SRE}\approx 1.47$~meV, see section~\ref{sec:Qdmodel}) and turn on LUT by varying the coupling strength by a factor $f_\text{LUT}$ between 0 and 1 in Fig.~\ref{fig:LUT}. As expected from the above considerations, the remaining twofold degeneracy in QD$^0$ is slightly lifted, causing a small splitting of the doublets. In charged QDs, the twofold degeneracy is not lifted. LUT contains interactions between HH and LH states. In our case of flat QDs, the lower excited transitions are mainly HH like and the occurring couplings are mainly between HH states due to a second order mechanism, mediated by LH states. An important effect of LUT is the reduction of the symmetry rules for couplings to $\Delta (\sum_{\text{particle},\alpha} a^{\text{particle}}_{\alpha}) \in \{0, 2, 4, ... \}$. Therefore LUT enables, within the considered energy range, the $p_y\to p_x$, $p_x\to p_y$, $p_z\to p_x$, $p_z\to p_y$ and $d_{xy}\to s$, $d_{xz}\to s$, $d_{yz}\to s$ as well as several $g\to s$ and LH $D\to s$ and $G\to s$ states to be coupled to the bright $s\to s$ or $p\to p$ states and therefore to become slightly bright. As a clear example one could follow the transition line appearing between the two solid blue $d_{xx}\to s$ and $d_{yy}\to s$ lines, namely the $d_{xy}\to s$ line in QD$^0$ or QD$^-$. In QD$^+$, the large number of lines prohibit an individual labeling, therefore we use colored areas to mark the average positions of the groups of peaks.

\subsection{Full model}\label{sec:Interactions:FullModel}
Within our full model, we consider DCI, SRE and LUT with appropriate values for CdSe ($\epsilon_r=9.2$, $M_\text{SRE}\approx 1.47$~meV, $f_\text{LUT}=1$). Thereby around half of the existing states are optically allowed in principle, although some will have a negligible oscillator strength. The other half, those with an uneven quantum number sum $\sum_{\text{particle},\alpha} a^{\text{particle}}_{\alpha}$, e.g. $p\to s$ transitions, will stay dark within our model, due to the assumed inversion symmetry.

The basic energetic splittings (Fig.~\ref{fig:INT}) within QD$^-$ and QD$^+$ let us presume a similar energetic structure for the trions. We find larger differences between p-shell transitions in QD$^-$ and QD$^+$. These differences are caused by mixtures between bright and nominally dark neighboring states. The extent of these mixtures strongly depends on the number of neighboring states as well as the energetic distance between these neighboring states and the bright p-shell transitions. Compared to QD$^-$, we find in QD$^+$ a larger number of nominally dark states as well as a smaller distance of these states to the p-shell transition lines. Thus the chance to find the p-shell transitions energetically clear cut with well defined spin configurations and high oscillator strength is much smaller in QD$^+$ than in QD$^-$.

\section{Geometry}\label{sec:Geometry}
In this section we study the influence of the geometry on the absorption. For this purpose, we vary the size, the in-plane asymmetry and the aspect ratio of the QD. We will use the full model.

\subsection{Size}\label{sec:Geometry:Volume}
In the following, we vary the QD size by a factor $l^3$ and fix the shape via ($l\cdot5.8$, $l\cdot5.0$, $l\cdot2.0$) nm$^3$.
The coupling strength scaling of the different interactions treated in this paper are known analytically for a HO confinement: By increasing the size, the single particle energies decrease by a factor $\sim l^{-2}$. Matrix elements of DCI have a smaller $\sim l^{-1}$ and matrix elements of SRE a larger $\sim l^{-3}$ dependency on $l$. In real systems, there may be slight deviations from these idealized dependencies: The single particle energies scale slower than $\sim l^{-2}$ due to band nonparabolicity effects \cite{franceschetti1997direct}. The dielectric constant in smaller QDs is reduced compared to its bulk value \cite{harvey1991raman, wang1994dielectric, laheld1997excitons, tsu1997simple}, leading to a larger DCI scaling than $\sim l^{-1}$. On the other hand, Ref. \Onlinecite{franceschetti1997direct} found a smaller scaling of DCI. The SRE scaling is found to be slower than $\sim l^{-3}$ (see Ref. \Onlinecite{tong2011theory}). However, here these deviations are supposed to be small enough to preserve the following results.

\begin{figure}
\includegraphics[width=0.45\textwidth]{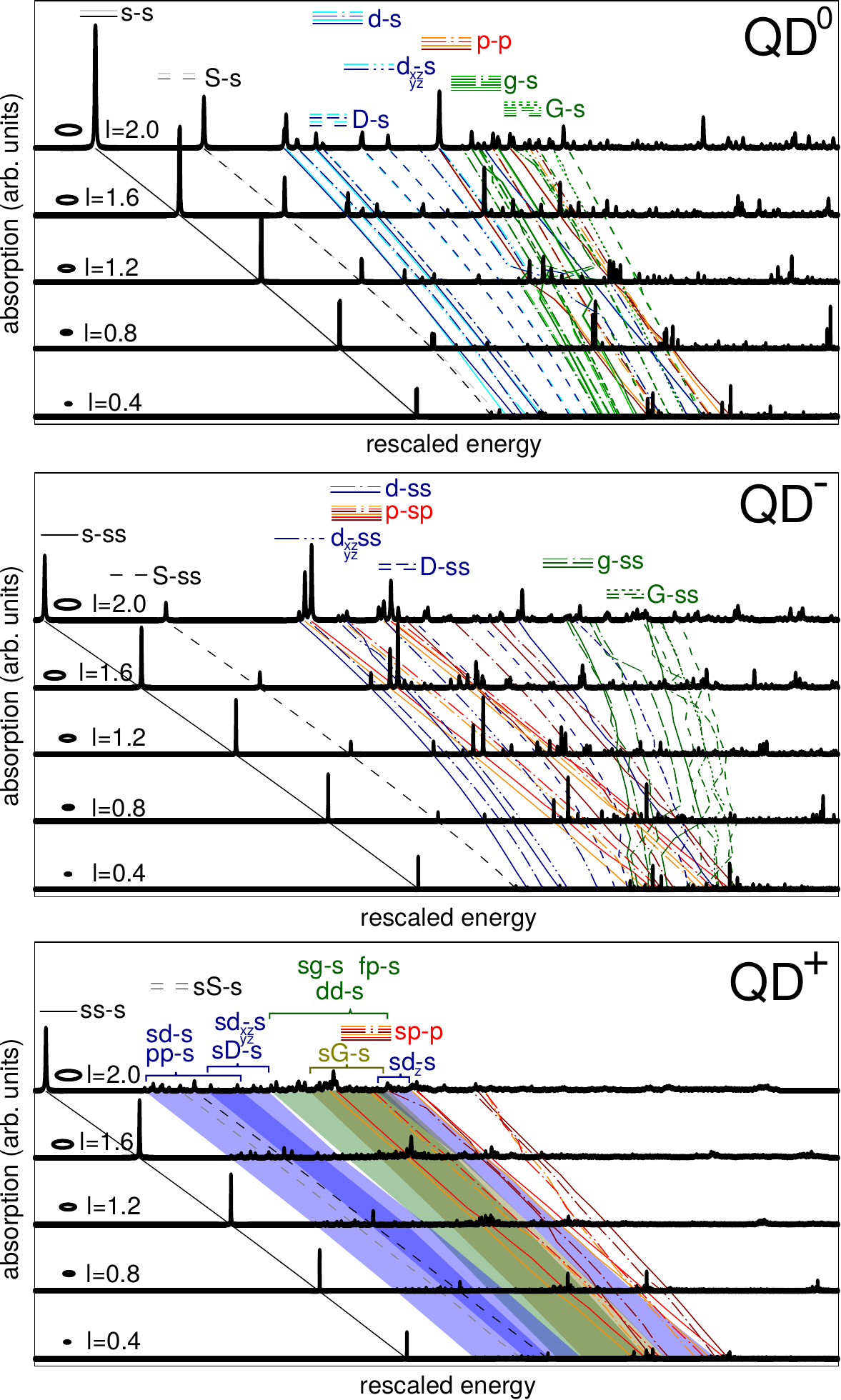}
\caption{(Color online) Absorption spectra for different QD sizes at fixed shape via ($l\cdot5.8$, $l\cdot5.0$, $l\cdot2.0$) nm$^3$. Energies (without bandgap) are rescaled by a factor $l^{-2}$. Colored lines and areas serve as a guide to the eye.}\label{fig:Volume}
\end{figure}

In larger QDs, a well known strong red shift of all states appears. The average distance between the levels decreases with the single particle energy spacings $\sim l^{-2}$. To visualize changes of the \textit{relative} energetic distances (e.g. whether $d\to s$ or $p\to p$ is closer to $s\to s$), we rescale the energies for each absorption spectrum by $l^{-2}$ and plot the spectra for different $l$ in Fig.~\ref{fig:Volume}. With increasing QD size, one observes a larger \textit{relative} singlet-triplet splitting in QD$^-$ and QD$^+$, an increasing \textit{relative} binding energy and the same \textit{relative} energy shifts and intensity changes as with increasing DCI (compare Figs.~\ref{fig:DCI} and \ref{fig:Volume}, most noticeable for QD$^-$). In fact, by increasing the QD size, we can directly affect the relative strength of DCI and therefore observe the same phenomena as with an increasing DCI coupling strength. This provides a good opportunity to study nearly pure impacts of DCI. Especially the correlations between $p-sp$ and $d-ss$, $g-ss$ in QD$^-$ can be tailored by changing the QD size. In QD$^+$ and QD$^0$ one observes more clear-cut and more intermixed $p\to p$ transitions, depending on the size, or more precisely on the question how exactly the different single particle energies of $g\to s$ and $p\to p$ match. These resonance effects in QD$^+$ and QD$^0$ on the absorption seem to be far beyond a possible technical control.

On the other side, the influence of SRE becomes relatively reduced in larger QDs (see coupling strength scaling). This is a much weaker effect and causes mainly a smaller relative distance between the two $p \to p$ triplet lines in larger charged QDs and a reduced relative bright-dark splitting in QD$^0$. As stated above, these effects of SRE might be even smaller than described here.

At this point, we discuss the often used classification of QDs into the strong and weak confinement, with respect to excited states. In a strong confinement, the Coulomb correlations become small and ultimately negligible compared to the subband energy spacings, thus the carriers can be described as single particles \cite{bryant1988excitons}. In contrast, the particles build a strongly correlated complex in the limit of the weak confinement.
Typically one considers the ground states for the definition of the strong and weak confinement and relates the cases descriptively with the ratio between QD diameter and bulk Bohr radius.
The ground states have in general a large subband energy spacing to the next higher states, whereas excited states are typically much closer to other states. Thus even if we can neglect the correlations of the ground states in good approximation and would define the QD to be in the strong confinement regime, it does not follow that we can neglect correlations for the excited states. This can also be seen in our calculations, where the ground states, labeled as $s\to s$, consist in small QDs ($l=0.4$) of around 95\% $s\to s$ states, whereas higher excited states like $p\to p$ ($d\to s$) consist of just around 56\% $p\to p$ (52\% $d\to s$). In larger QDs ($l=2.0$) these ratios decrease as expected and correlations become stronger. In the $s\to s$ we find just around 76\% and in the $p\to p$ ($d\to s$) around 38\% (33\%) of the respectively labeled states.

\subsection{In-plane Asymmetry}\label{sec:Geometry:InplaneAsymm}

\begin{figure}
\includegraphics[width=0.45\textwidth]{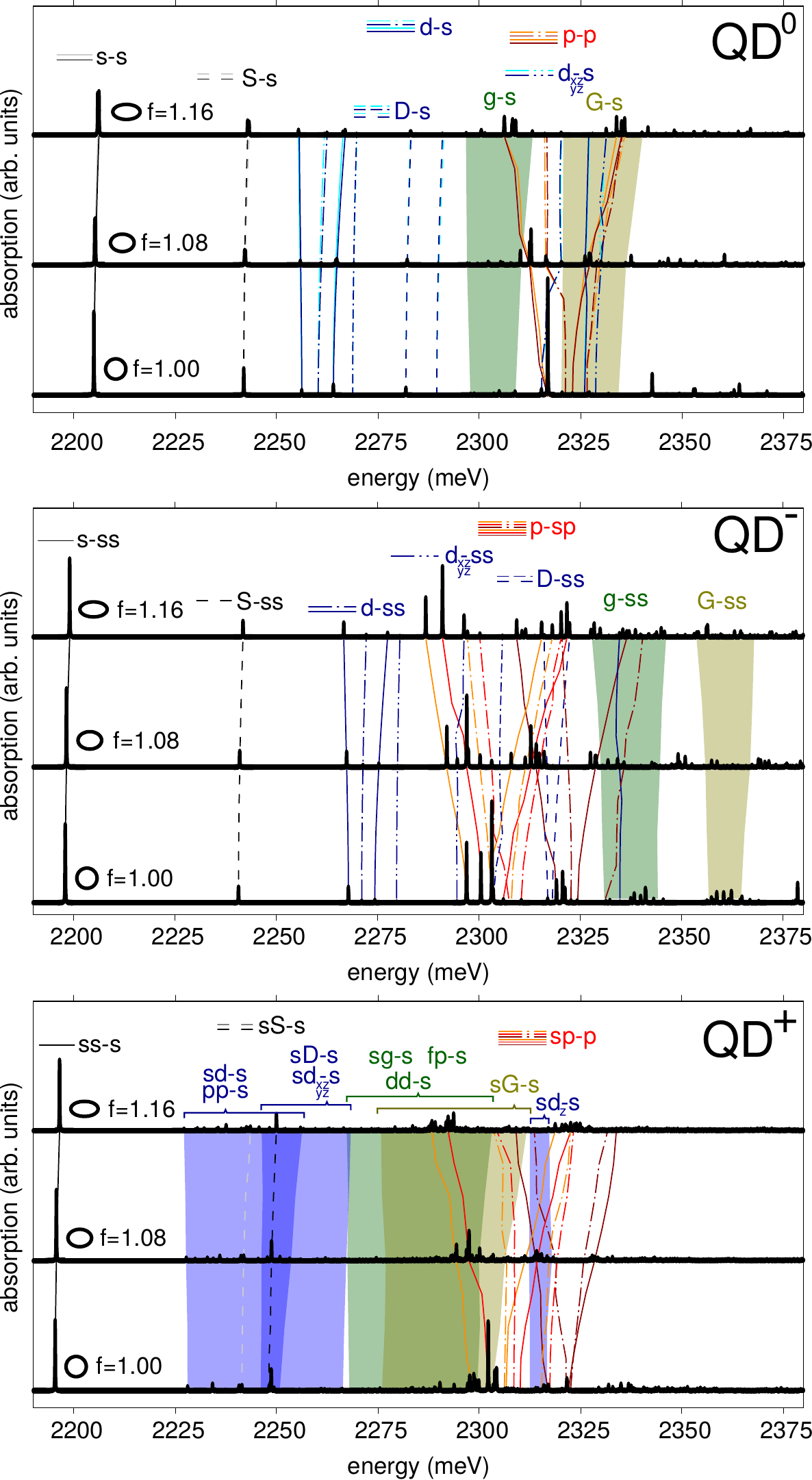}
\caption{(Color online) Absorption spectra for different in-plane asymmetry at fixed QD size and aspect ratio via ($\sqrt{f}\cdot\sqrt{5.8\cdot 5.0}$, $\frac{1}{\sqrt{f}}\cdot\sqrt{5.8\cdot 5.0}$, $2.0$) nm$^3$. Colored lines and areas serve as a guide to the eye.}\label{fig:InplaneAsymm}
\end{figure}

We study the influence of the in-plane asymmetry by fixing the size and aspect ratio of the QD and change the in-plane diameters via ($\sqrt{f}\cdot\sqrt{5.8\cdot 5.0}$, $\frac{1}{\sqrt{f}}\cdot\sqrt{5.8\cdot 5.0}$, $2.0$) nm$^3$. The previously used standard value for the QD in-plane asymmetry has been $f=1.16$. Absorption spectra for different asymmetry parameters $f$ are shown in Fig.~\ref{fig:InplaneAsymm}.

With increasing asymmetry, the main observation in all QDs is an increasing energetic separation between some lines, e.g. between $d_{xx}\to s$ and $d_{yy}\to s$ or between the different $p_x \to p_x$ and $p_y \to p_y$ lines (as an example, follow the $p_{x/y}-p_{x/y}$ lines in QD$^0$). This separation is caused by the increasing difference between the confinement lengths in $x$ and $y$ direction, leading to a larger separation of the single particle energies of states excited in different in-plane directions (SEDID). Without correlations, the energies of SEDID would meet at zero asymmetry. However, mainly DCI causes a coupling between those SEDID, resulting in one bright and one dark state (in our example $(p_x-p_x) \pm (p_y-p_y)$) at lower / higher energy. With increasing in-plane asymmetry, the energetic distance between the single particle energies increases, reducing the effective coupling until, at large asymmetries, the SEDID are mainly uncoupled (in our example uncoupled $p_x-p_x$ and $p_y-p_y$) and have nearly the same absorption intensity.

Another noticeable feature is the different gradient of the different SEDID, e.g. there is a much smaller separation of the $d\to s$ than of the $p\to p$ SEDID in all QDs. To clarify the origin of this finding, one should consider the single particle energy distance between SEDID $\sim (\frac{\Delta_e}{m_e}+\frac{\Delta_h}{\beta^2 m_h})(\frac{1}{L_x^2}-\frac{1}{L_y^2})$ with $\Delta_e$ ($\Delta_h$) being the envelope quantum number difference $a_x-a_y$ of the electron (hole) in $x$ / $y$ direction and $m_e$ ($m_h$) the effective masses in in-plane direction. Because $m_e$ is around three times smaller than in-plane $m_h$, $p \to p$ SEDID shift faster than $d \to s$ SEDID. Higher excited hole states (like $g \to s$ compared to $d \to s$) have a faster splitting caused by the larger $\Delta_e$/$\Delta_h$.

Besides, an overall slight blue shift in higher asymmetries is observed in all QDs. This is caused by the $\sim\frac{1}{L^2}$ dependency of the single particle energies on the confinement length $L$. Thus states excited in the direction of the narrower confinement (here $y$ direction) have a larger energetic increase with a confinement length reduction, than the energetic decrease of states excited in the direction of the broader confinement (here $x$ direction).

SRE effects seem to be stable related to anisotropy changes.

There are two possible profits from these findings:

1. The relative shifts between $p\to p$ and $d\to s$, $g\to s$ transitions in all QD charges enable a broad control of the mixtures and correlation strengths and the energetic order of several transitions. 

2. In principle, it is possible to determine the otherwise hard to assess in-plane asymmetry of a QD via the splitting of any particular pair of SEDID, measured e.g. in PLE. In practice, the single particle energies are influenced by several interactions, here mainly the QD confinement and DCI, thus sophisticated calculations are necessary to gain information about the asymmetry. A rough estimate of the asymmetry might be possible by comparing experimental data with Figs.~\ref{fig:Volume}, \ref{fig:InplaneAsymm} and \ref{fig:Height}. Fortunately, the most prominent and easy to identify peaks are the $p_x\to p_x$, $p_y\to p_y$ transitions, that undergo very similar effects under DCI as $s\to s$ (see binding energies in Fig.~\ref{fig:DCI}), what allows us to extinguish the influences of DCI in good approximation. Therefore, we can propose a very easy formula, deduces from single particle energies in the QD confinement, to determine the in-plane asymmetry parameter $f$ just via the energetic distance between the energy of the $s\to s$ transition ($E_{s\to s}$) and the different $p\to p$ transitions ($E_{p_{x/y}\to p_{x/y}}$) via
$$f=\frac{L_{x}}{L_{y}}\approx \sqrt{\frac{E_{p_y\to p_y}-E_{s\to s}}{E_{p_x\to p_x}-E_{s\to s}}}.$$
For very small asymmetries ($f \lesssim 1.04$), this method can not be used because the above described interaction via DCI between $p_x\to p_x$ and $p_y\to p_y$ at small asymmetries introduces strong deviations from the single particle energies and shade away the visibility of the $p_y\to p_y$ lines. For very large asymmetries ($f \gtrsim 1.8$), the energy spacing between $p_x\to p_x$ and $p_y\to p_y$ is so large, that correlations to the very different surrounding states lead to larger deviations ($\gtrsim 10\%$) in the prediction of $f$. For intermediate asymmetries, the differences between the $f$, used in our full model (see Fig.~\ref{fig:InplaneAsymm}) and predictions by the above easy formula are in the region of just a few percent. The above formula is independent from the QD size, aspect ratio, $\beta$, material or charge. We note, that strain effects might reduce the accurateness of the formula. In charged QDs, there exists no other method to determine the asymmetry via optical spectra, to our knowledge. In neutral QDs, the FSS of the bright ground states is caused by in-plane asymmetry, as stated above. Thus FSS measurements could in general reveal the asymmetry, especially useful to find QDs with nearly zero asymmetry \cite{takagahara1993effects,karlsson2010fine}. However, FSS seems to depend crucially on the coupling parameters of valence band mixing \cite{grundmann1995inas,seguin2005size} or long range Coulomb exchange \cite{goupalov1998anisotropic,takagahara2000theory}, thus on the material, strain, size, $\beta$ and probably on the charge, making this method more complicated.

\subsection{Aspect ratio}\label{sec:Geometry:Height}

\begin{figure}
\includegraphics[width=0.45\textwidth]{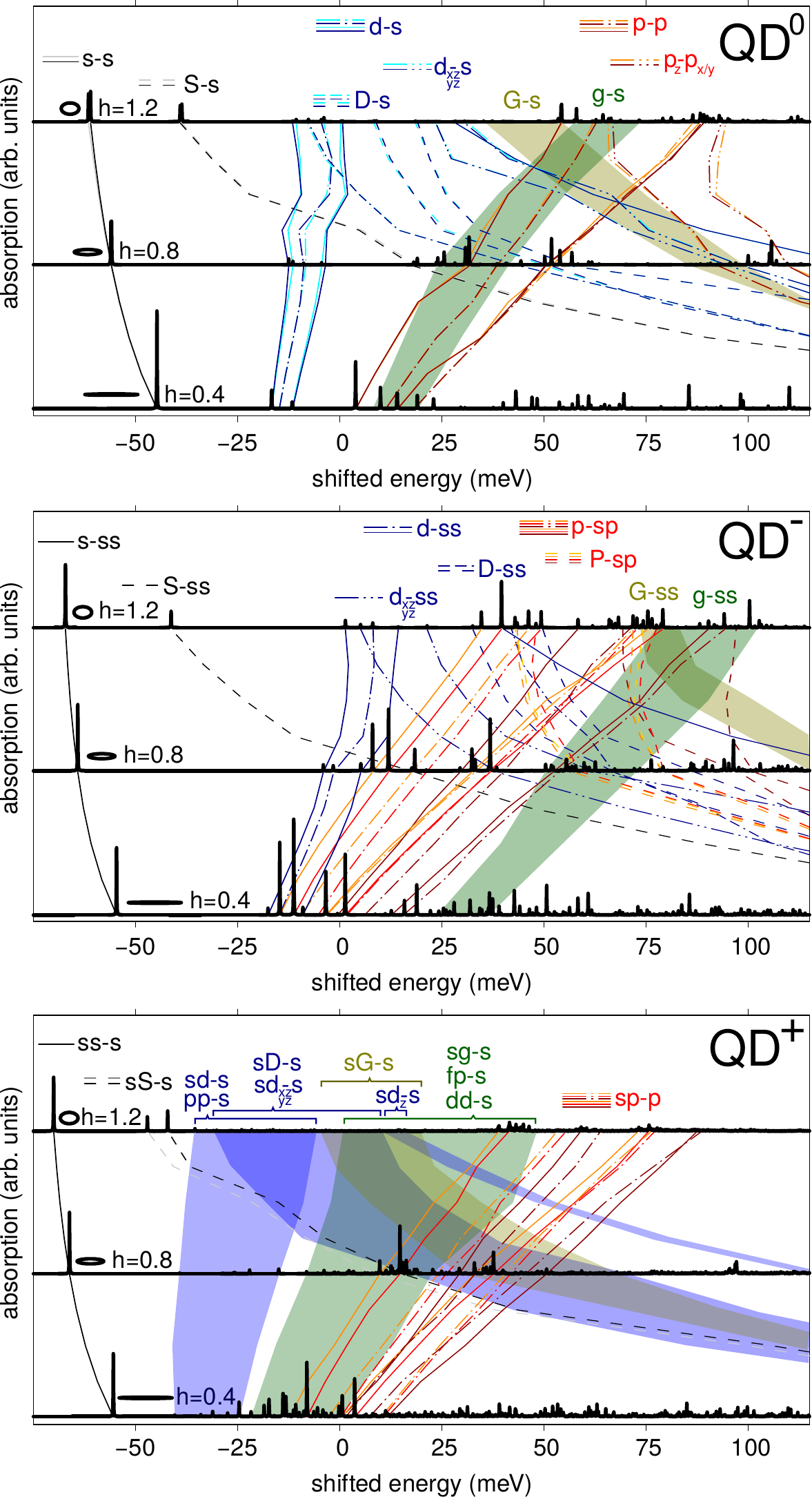}
\caption{(Color online) Absorption spectra for different QD aspect ratios at fixed in-plane asymmetry and size via ($\frac{1}{\sqrt{h}}\cdot5.8$, $\frac{1}{\sqrt{h}}\cdot5.0$, $h\cdot2.0$) nm$^3$. Energy is shifted respectively EMA energies. Colored lines and areas serve as a guide to the eye.}\label{fig:Height}
\end{figure}

In the following we study the changes in the electronic system caused by different QD aspect ratios. Therefore, we fix the QD size and in-plane asymmetry and vary the aspect ratio via ($\frac{1}{\sqrt{h}}\cdot5.8$, $\frac{1}{\sqrt{h}}\cdot5.0$, $h\cdot2.0$) nm$^3$. In flat QDs, the energy contribution of the confinement in $z$ direction dominates. Those contributions cause an overall strong red shift at larger $h$. To visualize the changes in level spacing and ordering, we shift (not rescale, as above!) the spectra by the single particle energies of the $s\to s$ transitions towards lower energies. Those shifted absorption spectra are plotted in Fig.~\ref{fig:Height} for different aspect ratio parameters $h$.

It can be seen that in higher QDs the LH states (short dashed lines) shift quickly towards lower energies, compared to the HH lines. In fact, if the QD height would be larger than its in-plane diameter, the LH $S\to s$ transition states would become the QD's ground states \cite{zielinski2013fine}. These shifts are caused by the different effective masses of LH and HH in $z$ direction: The energetic contribution in $z$ direction is $\sim\frac{1}{m_\text{LH/HH}^z}$. With $\frac{1}{m_\text{LH}^z} > \frac{1}{m_\text{HH}^z}$ the LH states have a larger dependency on the confinement length in $z$ direction, thus a faster red shift in higher QDs. In the shifted spectra just the relative red shift of the LH transitions compared to the HH transitions is visible.

A stronger red shift with increasing $h$ is also visible in states excited in $z$ direction (like the $d_z \to s$ states, right solid blue line), where the dominant term in $z$ direction is larger than in the other HH states, caused by the excitation in $z$ direction.

These different relative shifts enable an additional mechanism to tailor the correlations and mixtures, in this case between LH and HH states or states excited in in-plane and $z$ direction. Especially the possibility to change the LH contribution and therefore the spin state of a certain HH level might be interesting to control relaxation processes or oscillator strengths.

Another visible effect is the larger separation between states with different excitations in in-plane directions with larger $h$, thus between most states discussed in this paper, like $s\to s$, $d_{xx/xy/yy}\to s$, $p_{x/y}\to p_{x/y}$ and $g_{xxxx/xxxy/xxyy/xyyy/yyyy}\to s$. To fix the size, we decreased the in-plane diameter in higher QDs and consequently enhanced the in-plane single particle energy contributions. This has a larger effect on e.g. the $p_{x/y} \to p_{x/y}$ states than on the $d_{x/y} \to s$ states, because of the smaller effective mass of the electron than of the hole. This effect of energetic spacing between different shells is in good agreement with measurements in Ref. \Onlinecite{kuklinski2011tuning}. Also the splitting between $p_x\to p_x$ and $p_y\to p_y$ becomes enhanced in higher QDs, preserving the proportion to the distance between excited and ground states.

\begin{figure}
\includegraphics[width=0.45\textwidth]{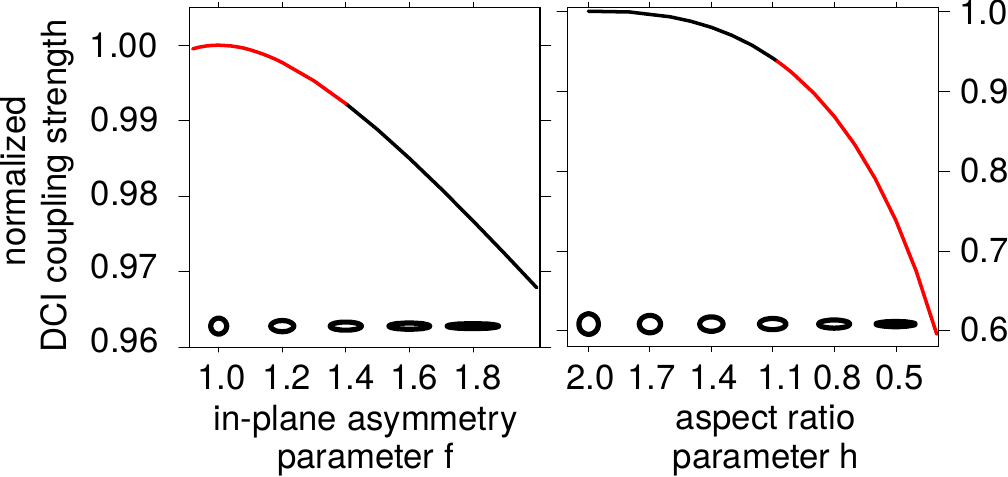}
\caption{(Color online) DCI matrix element between ground state heavy holes for different in-plane asymmetry parameters and aspect ratio parameters. Note different scales. Red areas depict typically relevant values for self assembled QDs.}\label{fig:CoulombIntAsymm}
\end{figure}

Finally, the Coulomb binding energy is enhanced with increasing $h$, visible e.g. in the stronger red shift of the $s\to s$ transitions. A closer analysis of DCI matrix elements in dependence on the deviation from the spherical shape shows the characteristics of a very broad Lorentzian function, centered at the spherical shape. The normalized DCI matrix elements for modifications of the aspect ratio and the in-plane asymmetry studied in this paper are plotted in Fig.~\ref{fig:CoulombIntAsymm}. All reasonable values for deviations from the spherical symmetry are closely around the tip of the Lorentz function. For asymmetries close to the sphere, like the in-plane asymmetry changes in section~\ref{sec:Geometry:InplaneAsymm}, the reduction of DCI is not noticeable. For the high differences between the QD elongation in in-plane and $z$ direction, that appear in this chapter, we get a strong effect, visible e.g. in the mentioned larger binding energies of the $s\to s$ transitions.

\section{Conclusion}\label{sec:Conclusion}

We have provided a detailed picture of the electronic energy structure of differently charged QDs focussed on p-shell transitions by studying different correlations and energetic trends in absorption spectra.
The individual and combined effects of DCI, SRE and LUT are described, highlighting the underlying processes behind energetic splittings and shifts as well as the reason for the appearance of additional lines in the absorption spectra and mixtures between different spin or spatial contributions to the absorption lines. Thereby we enable future studies of relaxation processes.
Our calculations predict larger differences in the absorption of negatively and positively charged QDs, where the chance to find clear-cut and well defined $p\to p$ transitions is larger in negatively charged QD than in positively charged QDs. We attribute these findings to a large number of nominally dark states close around the bright p-shell transitions in positively charged QDs.
We further studied changes of the absorption spectra with a modification of the QD's size and shape. Thereby we provide the knowledge to tailor the energetic structure, spin or spatial configuration or optical activity of excited states in a wide range. We clarify the classification of the strong and weak confinement regime with respect to excited states. We describe a simple method to gain information about the QD asymmetry from its absorption spectrum.

\begin{acknowledgments}
The authors thank Christopher Hinz, Denis Seletsky und Alfred Leitenstorfer for fruitful discussions.
\end{acknowledgments}

\appendix

\section{Phase convention}\label{sec:phase}
In the literature, there are several definitions of the relative phases between the Bloch functions. With LUT and SRE we use two interactions that mix Bloch states, thus we have to take special care of a consistent definition. The Bloch functions used in this paper are defined in terms of the spin states $\upharpoonleft$ and $\downharpoonleft$ as well as the spherical harmonics $Y_l^m$ in Condon-Shortley phase convention or the real valued cubic harmonics $S$, $P_x$, $P_y$ and $P_z$ via:
{\small{\begin{align*}
\edown   & = Y_0^0 \downharpoonleft                                                                 && = S \downharpoonleft  \notag\\
\eup     & = Y_0^0 \upharpoonleft                                                                   && = S \upharpoonleft    \notag\\
\HHdown  & = Y_1^1 \upharpoonleft                                                                   && =     - \sqrt{\frac{1}{2}} (P_x+iP_y) \upharpoonleft    \notag\\
\LHdown  & = \sqrt{\frac{1}{3}} Y_1^1 \downharpoonleft  + \sqrt{\frac{2}{3}} Y_1^0 \upharpoonleft   && =     - \sqrt{\frac{1}{6}} (P_x+iP_y) \downharpoonleft + \sqrt{\frac{2}{3}} P_z \upharpoonleft   \notag\\
\LHup    & = \sqrt{\frac{1}{3}} Y_1^{-1} \upharpoonleft + \sqrt{\frac{2}{3}} Y_1^0 \downharpoonleft && = \quad \sqrt{\frac{1}{6}} (P_x-iP_y) \upharpoonleft   + \sqrt{\frac{2}{3}} P_z \downharpoonleft \notag\\
\HHup    & = Y_1^{-1} \downharpoonleft                                                              && = \quad \sqrt{\frac{1}{2}} (P_x-iP_y) \downharpoonleft  \notag\\
\end{align*}}}


\end{document}